\documentstyle[12pt,epsf]{article}
\def\be{\begin{equation}}
\def\ee{\end{equation}}
\def\bea{\begin{eqnarray}}
\def\eea{\end{eqnarray}}

\begin{document}
\begin{titlepage}
\scrollmode
\vspace*{1cm}
\begin{center}
$$\;$$
{\large\bf  QUANTUM PARABOLIC SOMBRERO}
\end{center}
\begin{center}
Ye. Hakobyan${}^1$,
S. Ter--Antonyan${}^2$,
V. Ter--Antonyan${}^3$
\\[3mm]

\footnotetext[1]{e-mail: yera@thsun1.jinr.ru}
\footnotetext[2]{e-mail: simonter@hotmail.com}
\footnotetext[3]{e-mail: terant@thsun1.jinr.ru}

\vspace{0.5cm}

{\small Bogoliubov Laboratory of Theoretical Physics} \\
{\small Joint Institute for Nuclear Research} \\
{\small Dubna, Moscow Region, 141980, Russia}
\end{center}

\vspace{1cm}

\begin{abstract}
We have discussed the energy levels and probability distribution
density for a quantum particle placed in the
two-dimensional sombrero-shaped potential
$V(\rho,\rho_0)=\mu\omega^2|\rho^2-\rho_0^2|/2$.
\end{abstract}

\vspace{0.7cm}


\end{titlepage}

\section{Introduction}

It is well known \cite{Ok} that potentials having the form of a sombrero
are convenient for explanation  of
spontaneous breaking  of continuous symmetry.

On the other hand, until now there is no quantitative analysis for the motion of
a quantum particle in this potential. The cause seems to be
in that the sombrero, as a smooth potential, is a polynomial of the
forth power, and for that potential the Schr\"oedinger equation is not
exactly solvable. However, there is no contradiction in
refusing the smoothness condition and constructing a
sombrero by means of the quadratic potential
$W(\rho,\rho_0)=\mu\omega^2(\rho-\rho_0)^2/2$, where $\rho=\sqrt{x^2+y^2}$ and
$\rho_0\in [0,\infty)$ is a parameter.
The one-dimensional version of the potential $W(\rho,\rho_0)$ is known
as a double-oscillator \cite{Mer}.
The smoothness is broken for
$W(\rho,\rho_0)$ at the point $\rho=0$.
Nevertheless, the potential
$W(\rho,\rho_0)$ is still complicated as the corresponding radial equation
includes along with the terms $\rho^2$ and $\rho^{-2}$ the linear term $\rho$.
This type of equations is not exactly  solvable. One can go further and break
the smoothness condition not at one but at an infinite number of points. The
simplest possibility is realized by the potential $V(\rho,\rho_0)= \mu \omega^2
|\rho^2-\rho_0^2|/2$, which we call the parabolic sombrero.  Here, the
smoothness is broken along the circle of the radius $\rho_0$. For $y=0$ (or
$x=0$) the parabolic sombrero transforms into a two-center oscillator considered
in paper \cite{1d}. The aim of the present article is to calculate
the energy spectrum and wave functions of a quantum particle placed in
the parabolic sombrero potential.

\section{Spectral Equation}

The Hamiltonian of the parabolic sombrero has the form
$$
H(\rho, \rho_0)=-\frac{\hbar^2}{2\mu}\left(
\frac{\partial^2}{\partial \rho^2}+\frac{1}{\rho}
\frac{\partial}{\partial \rho}
+\frac{1}{\rho^2} \frac{\partial^2}{\partial \varphi^2}
\right)+
\frac{\mu \omega^2}{2} |\rho^2-\rho_0^2|,
$$
where $\rho$ and $\phi$ are polar coordinates of the particle:
$0\leq\rho<\infty$, $0\leq\varphi< 2\pi$.
To this Hamiltonian there correspond two radial equations
$$
\frac{d^2 R_{in}}{d r^2}+\frac{1}{r}
\frac{dR_{in}}{d r}
+\left(\frac{r^2}{4}-\frac{m^2}{r^2}- \xi_{in}\right)R_{in}=0,
$$
$$
\frac{d^2 R_{out}}{d r^2}+\frac{1}{r}
\frac{dR_{out}}{d r}
-\left(\frac{r^2}{4}+\frac{m^2}{r^2}+ \xi_{out}\right)R_{out}=0,
$$
where $r=(2\mu\omega/\hbar)^{1/2}\rho,
r_0=(2\mu\omega/\hbar)^{1/2}\rho_0,
\varepsilon=E/(\hbar\omega)$ and $m\in {\bf Z}$ is the eigenvalue
of the angular momentum operator $\hat L=-i \partial/\partial \varphi$,
$$
\xi_{in}=r_0^2/4-\varepsilon,\quad
\xi_{out}=-r_0^2/4-\varepsilon.
$$
The first equation is for the inner region $[0,r_0)$ of the sombrero,
while the second one is for its outer $(r_0,\infty)$ region.
We are interested in the solutions finite at $r=0$ and vanishing as
$r\to\infty$. These conditions select the functions
$$
R_{in}(r)=C_{in}
r^{|m|} e^{-i r^2/4}
F(\alpha, \gamma;ir^2/2),
$$
$$
R_{out}(r)= C_{out} \,  r^{|m|} e^{-r^2/4}
\Psi(a,\gamma;r^2/2).
$$
Here $F$ and $\Psi$ are two independent solutions of the confluent
hypergeometric equation:
$$
F(\alpha,\gamma; z)=1+ \frac{\alpha}{\gamma}\frac{z}{1!}+
\frac{\alpha(\alpha+1)}{\gamma(\gamma+1)}\frac{z^2}{2!}+\dots,
$$
$$
\Gamma(a)\Psi(a,b;z)=
\int\limits_{0}^{\infty} e^{-z t}t^{a-1}
(1+t)^{b-a-1} dt,
$$
$$
\alpha=\frac{|m|+1- i \xi_{in}}{2},\quad
a=\frac{|m|+1+\xi_{out}}{2},\quad \gamma=|m|+1,
$$
and $C_{in}$ and $C_{out}$ are normalization constants.
The formulae given below from the theory of confluent
hypergeometric equation are taken from the monograph \cite{Bat}.

Let us require the equality of the logarithmic derivatives of the functions
$R_{in}$ and $R_{out}$ at the point $r=r_0$.
This condition works for
$0< r_0 < \infty$. For $r_0=0$
the smoothness at $r=r_0$ is observed just
for $R_{in}$ and $R_{out}$, but
not  their derivatives (see for example \cite{Land}).
After using formulae
$$
F'(\alpha, \gamma; z)=\frac{\alpha}{\gamma} F(\alpha+1, \gamma+1; z),
$$
$$
\Psi'(\alpha, \gamma; z)=-a \Psi(a+1, \gamma+1; z),
$$
we come to the spectral equation
$$
\frac{i \alpha}{\gamma}\,\,\,
\frac{F\left(\alpha+1,\gamma+1;i z_0\right) }
{F\left(\alpha,\gamma;i z_0\right) }
-\frac{i}{2}= -a \,\,\,\frac{\Psi\left(a+1,\gamma+1;z_0\right)}
{\Psi\left(a,\gamma;z_0\right)}-\frac12,
$$
where $z_0=r_0^2/2$.

Note that from the Kummer transformation
$$
F(\alpha, \gamma; z)=e^z F( \gamma-\alpha, \gamma; -z),
$$
the relation $\gamma-\alpha^*=\alpha$ and the recurrent formula
$$
(\alpha-\gamma) F(\alpha, \gamma+1;i r_0^2/2)=
-\gamma F(\alpha, \gamma;i r_0^2/2) +\alpha
F(\alpha+1,\gamma+1;i r_0^2/2)
$$
there follows that the left
hand side of the spectral equation is real.
From the Kummer
transformation there follows that the function $R_{in}$ is also real.

In the next section, we will discuss the results obtained from the spectral
equation by numerical calculations.

\section{Energy Levels}

For $r_0=0$ the parabolic sombrero transforms into the circular oscillator for
which the state is determined by the quantum numbers $(n_r, m)$, and the $n$-th
energy level is given by the formula $\varepsilon=n+1$ and has multiplicity
$g_n=n+1$ ($n_r$ is the number of zeroes of the radial wave function in
the region $(0,\infty)$, $n=2 n_r+|m|$).

The inclusion of the parameter $r_0$ splits the energy levels
and transforms them into an infinite set of intersecting lines in the plane
$(\varepsilon, r_0)$ composing a complicated picture (see Fig. 1).

Let us separate the energy lines of the parabolic sombrero into clusters of
three types: with fixed $n$ ($n$-cluster), $|m|$ ($|m|$-cluster)
and $n_r$ ($n_r$-cluster).

\begin{figure}[htb]
\vspace*{-1cm}
\centerline{
\vbox{
\epsfysize=90mm
\epsfbox{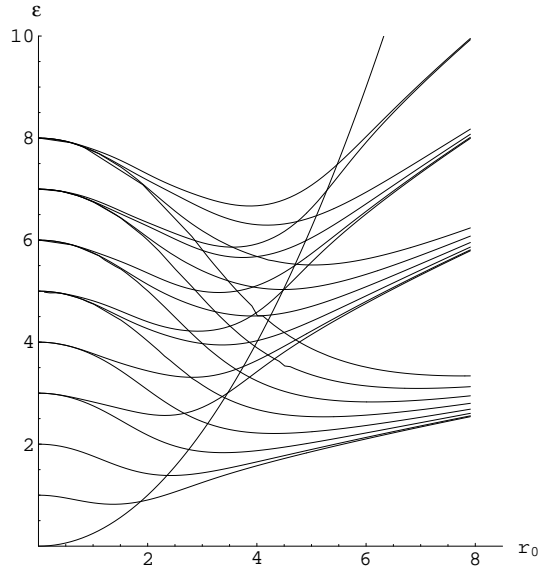}
}
}
\caption{ Dependence of the energy levels of the parabolic
sombrero on $r_0$. Here as well as in
Figs. 2--5 the parabola describes  the parabolic sombrero
potential as a function of $r_0$ for $r=0$.}
\label{fig:1}
\end{figure}


{\bf a.} The $n$-cluster possesses $(n/2+1)$ or $(n+1)/2$
lines for even and odd $n$, respectively (see Fig. 2).
For large values of the parameter
$r_0$, the lines of the $n$-cluster are very much separated
from each other.
With decreasing parameter $r_0$ the distance between
the lines of the $n$-cluster decreases.
Beginning at some $r_0$, particular for every line,
the lines jump above the top of the parabolic sombrero
and with the following decrease of $r_0$ the lines combine in one
$(n+1)$ degenerate energy level of a circular oscillator.


\begin{figure}[htb]
\centerline{
\vbox{
\epsfysize=80mm
\epsfbox{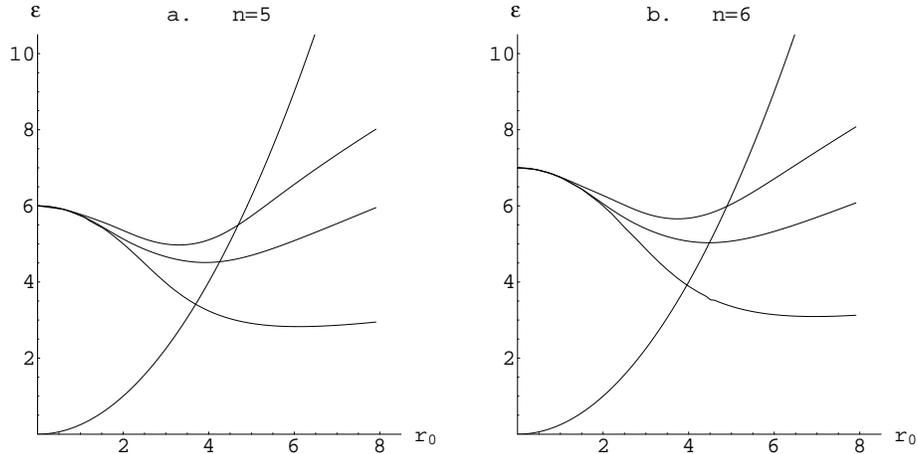}
}
}
\caption{
Dependence of the lines of $n$-clusters on the parameter $r_0$
for $n=5$ and $n=6$.  To larger values of $\varepsilon(r_0)$
(for fixed $r_0$) corresponds  the larger $n_r$.}
\label{fig:2}
\end{figure}


{\bf b.} Every $|m|$-cluster includes an infinite number of lines (see Fig. 3).
With growth of the parameter $r_0$, each line of the $|m|$-cluster,
located above the top of the sombrero, first slightly lowers, then grows
and starting from some
parameter $r_0$, particular for each line,
is captured  by a circular wall. For further growth of $r_0$
the lines of the $|m|$-cluster grow with different velocity: the more $n_r$
the faster the growth of the line.


\begin{figure}[htb]
\centerline{
\vbox{
\epsfysize=125mm
\epsfbox{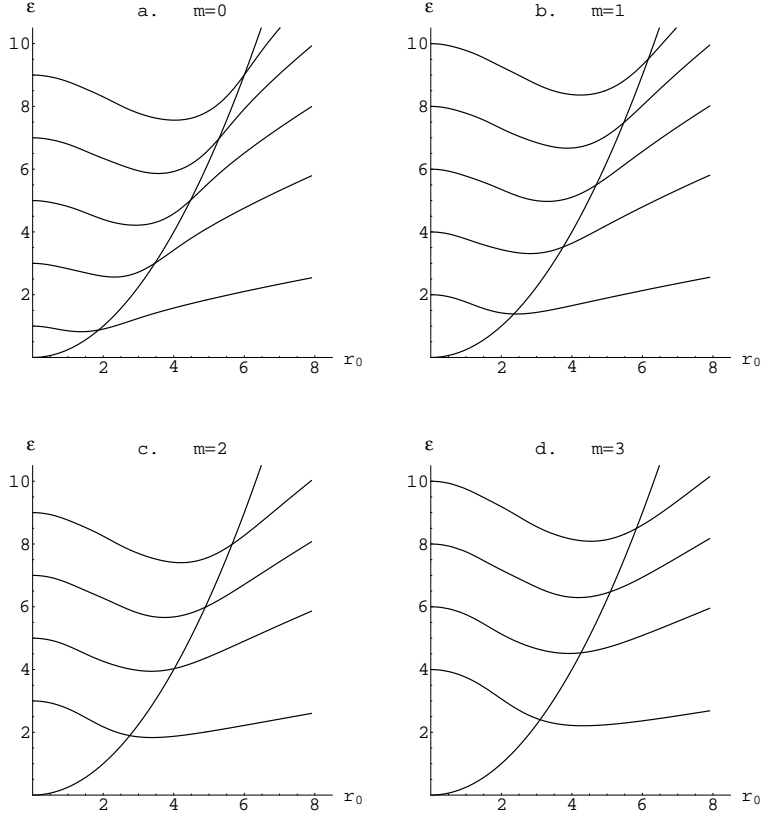}
}
}
\caption{
Dependence of the lines of $|m|$-clusters on the parameter $r_0$
for $m=0, 1, 2, 3$.}
\label{fig:3}
\end{figure}


{\bf c.} Let us compare the lines of $|m|$-cluster with $m=0$ and
the energy levels of the two-center quantum oscillator \cite{1d}.
The equation describing the two-center oscillator can not be obtained from
radial equation of the parabolic sombrero: if we substitute $m=0$
and $R= f/r^{1/2}$ into the radial equation, we eliminate the
centrifugal potential $m^2/r^2$ and delete the term with
first derivative, but then arises an additional centrifugal potential.
This new centrifugal potential influences the energy spectrum, as
the result of which the spectroscopy of the parabolic sombrero with $m=0$
and the spectroscopy of the two-center oscillator are not identical.
As we can see in Fig. 4, with the increase in the parameter $r_0$
the energy levels of the two-center quantum oscillator merge in pairs
and between them is the line corresponding to the parabolic sombrero.


\begin{figure}[htb]
\centerline{
\vbox{
\epsfysize=90mm
\epsfbox{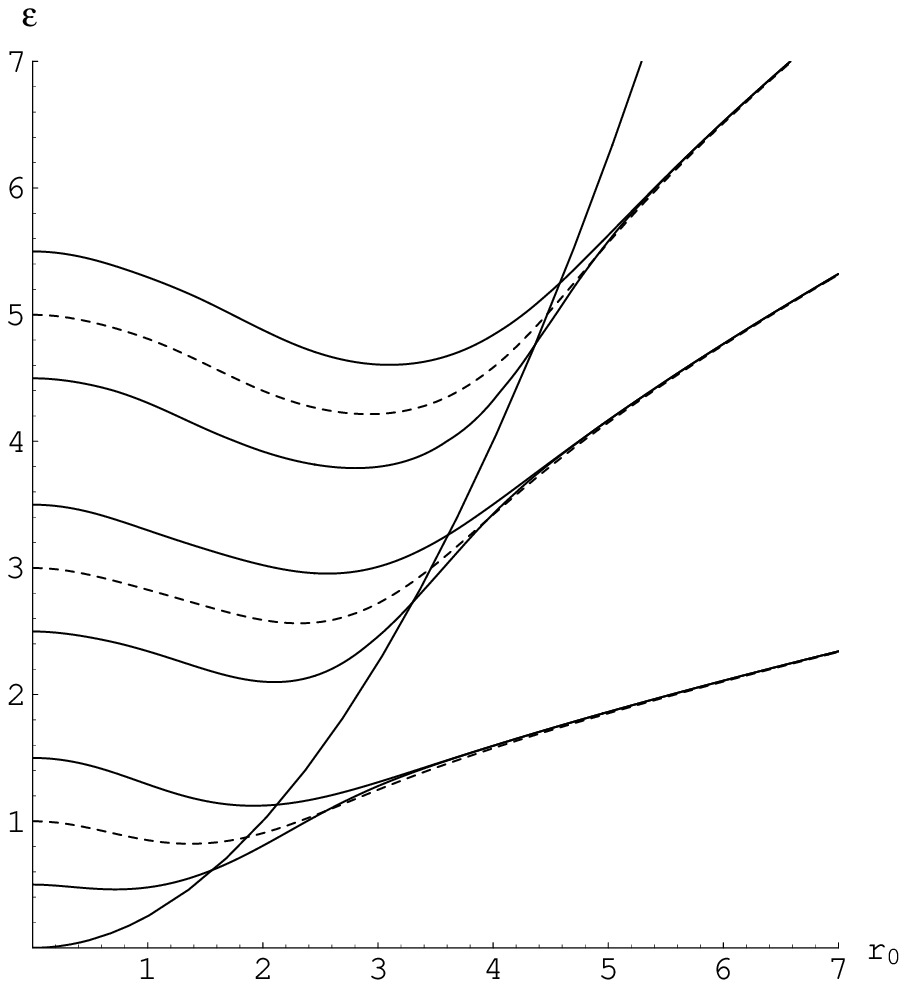}
}
}
\caption{ Comparison of the
lines of the "$m=0$"-cluster of the parabolic sombrero with the
energy levels of the two-center quantum oscillator taken from \cite{1d}.
Solid lines correspond to the two-center quantum oscillator;
dotted lines to the parabolic sombrero.}
\label{fig:4}
\end{figure}


{\bf d.} Quite interesting is the behavior of the lines of $n_r$-clusters
(see Fig. 5).  Every $|m|$-cluster, as well as $n_r$-cluster, includes an
infinite number of lines.
With the growth of the parameter $r_0$, the lines
of the $n_r$-cluster are gradually coming together,
then captured by the circular wall,
and continuing the approach merge into one line.
Thus, for $r_0\to\infty$ we have an infinite number of levels,
each being an infinite degenerated $n_r$-cluster.


\begin{figure}[htb]
\centerline{
\vbox{
\epsfysize=80mm
\epsfbox{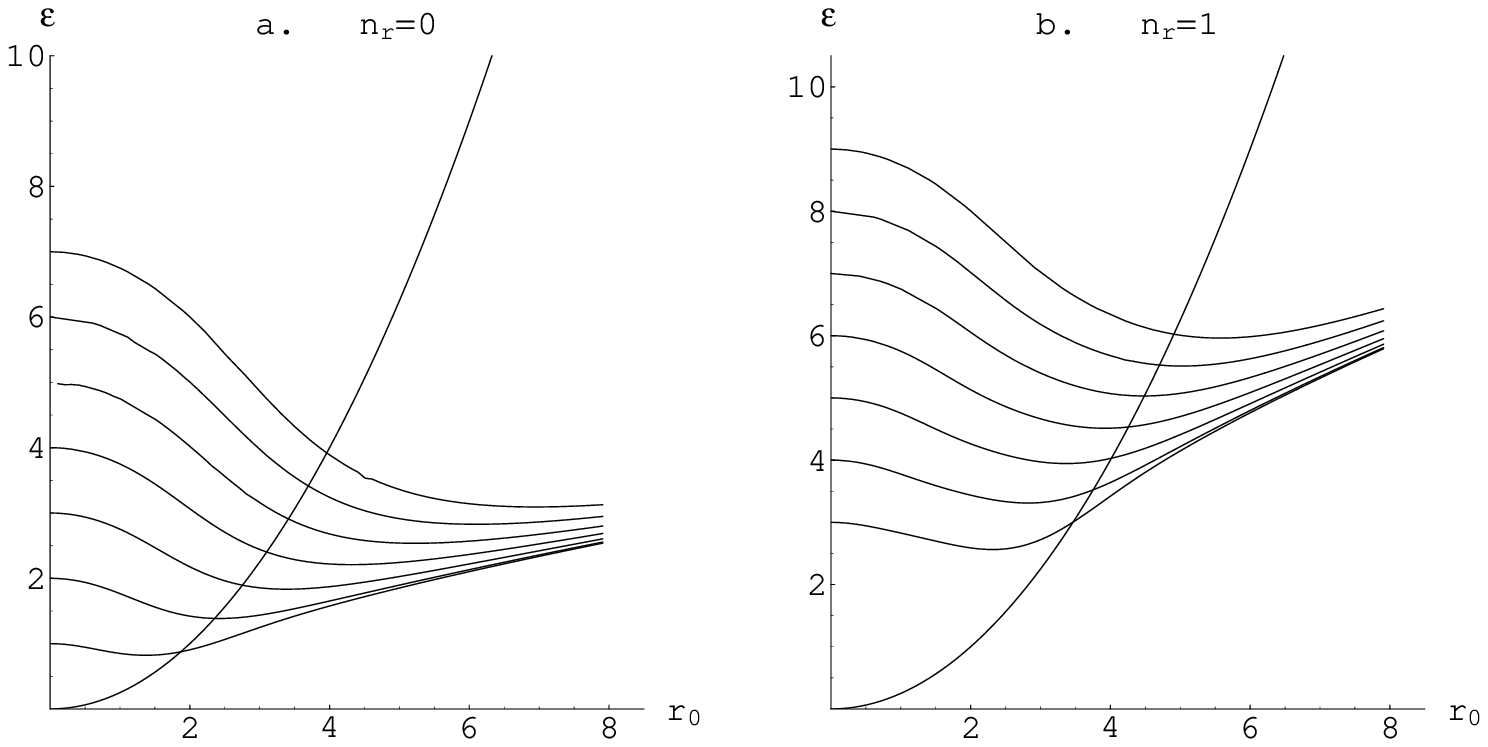}
}
}
\caption{
Dependence of the lines of $n_r$-clusters of the parabolic sombrero on
the parameter $r_0$ for $n_r= 0, 1$.}
\label{fig:5}
\end{figure}


{\bf e.}
Using the known formula \cite{Land}
$$
\frac{\partial E_{n_r,m}}{\partial \rho_0}=
\left(
\frac{\partial \hat{H}}{\partial \rho_0}
\right)_{n_r,m;n_r,m},
$$
we obtain
$$
\frac{\partial E_{n_r,m}}{\partial \rho_0}=
{\mu\omega^2} \rho_0\left[
2 \int\limits_{0}^{\rho_0}
(R_{n_r,m}^{in})^2 \rho \, d\rho -1
\right].
$$
The integral in the brackets is a monotonously increasing function of
$\rho_0$ with  the range of values $[0,1]$. Expanding it in powers of
$\rho_0$ (for small $\rho_0$) and $1/\rho_0$ (for large $\rho_0$), we obtain
$$
\varepsilon_{n_r,m}(r_0)\simeq
2 n_r+|m|+1-\frac{r_0^2}{4},
\quad \mbox{for}
\quad r_0\ll 1,
$$
$$
\varepsilon_{n_r,m}(r_0)\simeq
\frac{r_0^2}{4} - A_{n_r,m} r_0,
\quad \mbox{for}
\quad r_0\gg 1,
$$
where $A_{n_r,m}$ is the matrix with positive elements independent of
$r_0$. As will be shown below (see Conclusion), for extremely large values of the
parameter $r_0$, the quantities $A_{n_r,m}$
cease to depend on the quantum number $m$.

We come to the following conclusions: firstly, the corrections to the
spectrum of the circular oscillator $(r_0\ll 1)$ are not linear but quadratic
in $r_0$; secondly, after the capture by the wall
the energy levels accumulate near the top of the sombrero forming a ring with
thickness $r_0$ times less than the top of the sombrero.
The last conclusion is confirmed by Fig. 5.

\section{Probability Distribution}

Let us introduce two functions
$$
D_{in}(r)=
r^{|m|} e^{-i r^2/4}
F(\alpha, \gamma;ir^2/2),
$$
$$
D_{out}(r)= r^{|m|} e^{-r^2/4}
\Psi(a, \gamma;r^2/2),
$$
and rewrite the functions  $R_{in}$ and $R_{out}$
in the form
$$
R_{in}(r)=C_{in}D_{in}(r),
\qquad
R_{out}(r)= C_{out}D_{out}(r).
$$
To find the normalization constants, we just sew together
$R_{in}$ and $R_{out}$ at the point $r=r_0$:
$$
C_{in}D_{in}(r)=
C_{out}D_{out}(r),
$$
and demand
$$
C_{in}^2\int\limits_{0}^{r_0} D_{in}^2 r \,dr+
C_{out}^2 \int\limits_{r_0}^{\infty} D_{out}^2 r \,dr = 1.
$$
It is easy to conclude from these two equations that
$$
C_{in}= D_{out}(r_0)/Q(r_0), \qquad
C_{out}= D_{in}(r_0)/Q(r_0),
$$
where
$$
Q(r_0)= \left[
D_{out}^2(r_0)
\int\limits_{0}^{r_0}D_{in}^2(r) r dr+
D_{in}^2(r_0)
\int\limits_{r_0}^{\infty}D_{out}^2(r) r dr
\right]^{\frac12}.
$$
The obtained formulae allow one to look at the picture of motion of the energy
levels from the point of view of probability distribution.
The diagrams of numerical calculations for the functions $r R^2$ for
$n_r=3$, $m=0$ and different values of the parameter $r_0$
are presented in Fig. 6.
Considering these diagrams in ascending order of the parameter $r_0$
we see that they confirm the general scenario of the motion of the levels
described in the previous section.


\begin{figure}[htb]
\centerline{
\vbox{
\epsfysize=100mm
\epsfbox{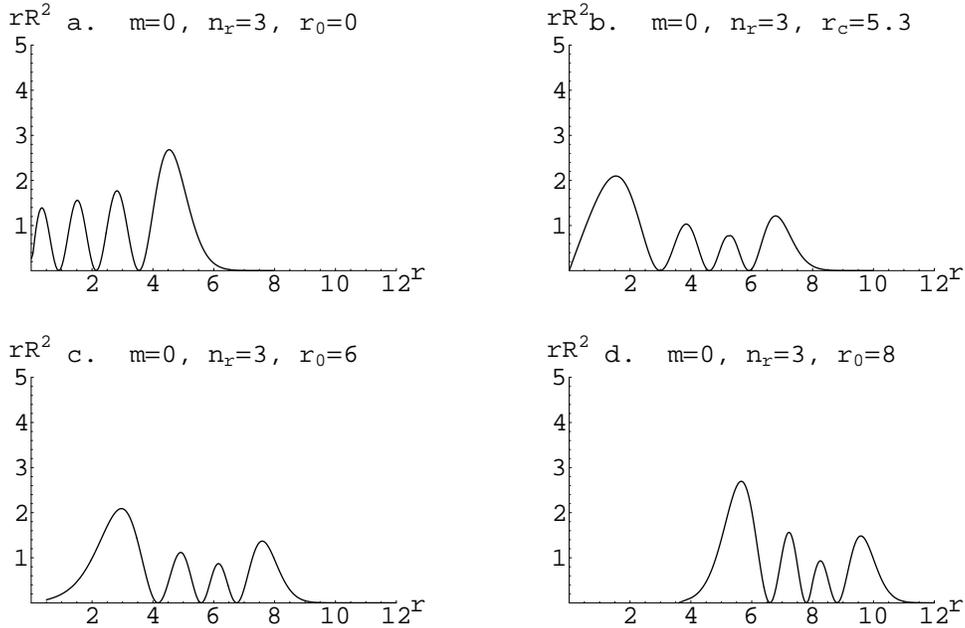}
}
}
\caption{
Diagrams of the
probability distributions for $m=0, n_r=3$ and different values of $r_0$.}
\label{fig:6}
\end{figure}


Fig. 6a corresponds to the circular oscillator.
In Fig. 6b  the moment of capture of the level by the wall ($r_c$ is
the value of the parameter $r_0$ when the capture takes place) is
drawn.  Figs. 6c and 6d demonstrate the shift of the
particle together with the wall far from the origin of the coordinates
with further increase in the parameter $r_0$.

\section{Conclusion}

In conclusion, let us discuss the limit $r_0\to\infty$
more thoroughly. As $r_0\to\infty$ the height of the barrier ($r_0^2/4$)
increases, and that is why the wave functions, corresponding to the
energy levels in the wall differ from zero only in the region of large $r$,
where the centrifugal term $m^2/r^2$ could be neglected. In such an approach
the functions $R_{in}$ and $R_{out}$ as well as the energy levels
cease depending on quantum number $m$, which in its turn indicates the
dependence (in the limit of large $r_0$) of the spectroscopy of captured
levels practically on the quantum number $n_r$ only.
Such a conclusion is in agreement with the
tendency, presented in Fig. 5,
and with the general philosophy of spontaneous breaking of
continuous global symmetry \cite{Ok}.

\section*{Acknowledgment}
The authors would like to thank G. Pogosyan and A. Sissakian for
fruitful discussions. The work of Ye. Hakobyan was partially
supported by Russian Foundation for Basic Research
(RFBR) under the project \# 98-01-00330.

\newpage


\end{document}